\documentclass[reprint, amsmath, amssymb, aps, prl]{revtex4-1}
\usepackage{graphicx}
\usepackage{amssymb}
\usepackage{amsmath}
\usepackage{dsfont}
\usepackage{setspace}
\usepackage{dcolumn}
\usepackage{newfloat}
\usepackage{epsfig}
\usepackage{multibib}

\usepackage[colorlinks=true, citecolor=blue, linkcolor=blue, urlcolor=blue]{hyperref}

\usepackage{epstopdf}
\usepackage{bm}
\usepackage{float}
\usepackage{color}

\def\bea{\begin{eqnarray}}
\def\eea{\end{eqnarray}}
\def\be{\begin{equation}}
\def\ee{\end{equation}}

\usepackage{datetime}
\advance\currenthour by 0

\begin{document}

\title{Emergent phases and novel critical behavior in a non-Markovian open quantum system}
\author{H. F. H. Cheung, Y. S. Patil and M. Vengalattore}
\affiliation{Laboratory of Atomic and Solid State Physics, Cornell University, Ithaca, NY 14853}
\email{mukundv@cornell.edu}

\begin{abstract}
Open quantum systems exhibit a range of novel out-of-equilibrium behavior due to the interplay between
coherent quantum dynamics and dissipation. Of particular interest in these systems are driven, dissipative transitions,
 the emergence of dynamical phases with novel broken symmetries, and critical behavior that lies
beyond the conventional paradigms of Landau-Ginzburg phenomenology. Here, we consider a 
parametrically driven two-mode system in the presence
of non-Markovian system-reservoir interactions. We show that non-Markovianity modifies the phase diagram of this system resulting in the emergence
of a novel broken symmetry phase in a new universality class that has no counterpart in a Markovian 
or equilibrium system. Such reservoir-engineered dynamical phases can potentially shed light on universal aspects
of dynamical phase transitions in a wide range of non-equilibrium systems, and aid in the development
of techniques for the robust generation of entanglement and quantum correlations at finite temperatures
with potential applications to quantum metrology. 
\end{abstract}

\maketitle
{\em Introduction.} Due to the commensurate influence of quantum coherence and
dissipation, the dynamical behavior of open quantum systems conforms neither to the framework of unitary quantum 
evolution nor to thermodynamic descriptions \cite{rivas2011}. Motivated
by various applications to quantum information science, experimental realizations of such open systems have been
developed in platforms spanning ultracold atomic gases \cite{schindler2012}, circuit-QED systems \cite{xiang2013}, optomechanical systems \cite{aspelmeyer2014} and hybrid
quantum systems \cite{kurizki2015}. The exploration of novel dynamical phases and the development of techniques for robust quantum state preparation and control in these systems 
presents significant theoretical and experimental challenges that lie at the interface of atomic physics, quantum optics, 
 and condensed matter physics. 

In addition to the traditional approach of Hamiltonian design, open quantum systems are amenable to control by modifying the nature of their 
environment. As such, the concept of reservoir-engineering \cite{diehl2008} has emerged as a promising paradigm for the realization
of novel states of open and driven quantum systems. In certain cases, it has been shown that reservoir-engineering can be used to coax the open 
quantum system into phases that might not be accessible through more conventional forms of quantum 
state preparation \cite{muller2012,tomadin2012}. 
Aside from presenting alternate routes to quantum state preparation, such reservoir-engineered quantum 
phases present intriguing questions in their own right. For instance, it is unclear to what extent driven, dissipative
transitions in open quantum systems accommodate the central paradigms of scale invariance, symmetry breaking and
universality that underpin our understanding of equilibrium and quantum phase transitions. 

Here, we explore the driven, dissipative transitions of a parametrically driven two-mode 
quantum system in the presence of a non-Markovian environment. This is a minimal physical realization of the 
parametric oscillator model \cite{drummond2005,wouters2007} and is closely connected to the open Dicke model \cite{emary2003, baumann2010,klinder2015}, the superradiant phase
transition \cite{dicke1954} and the Lipkin-Meshkov-Glick model \cite{lipkin1965}. 
In the presence of a Markovian reservoir, this system exhibits a non-equilibrium phase transition into an ordered
state that develops beyond a critical magnitude of the external drive \cite{dechoum2004}. 
Going beyond the Markovian regime, recent work has shown that the presence of a sub-ohmic reservoir modifies the critical
exponents of this non-equilibrium transition while preserving the steady-state phase diagram \cite{dallatorre2010,nagy2011,nagy2016}. 
In this work, we identify a class of experimentally accessible non-Markovianity that leads to significant changes in the phase
diagram of this system, leading to the emergence of a dynamical phase with novel broken symmetries and critical behavior
that is distinct from that observed in a Markovian system. 
 We demonstrate that the novel emergent phase manifests significantly enhanced correlations and entanglement than 
 can be realized in a Markovian system. 
 
 \begin{figure}[t]
\includegraphics[width=0.4\textwidth]{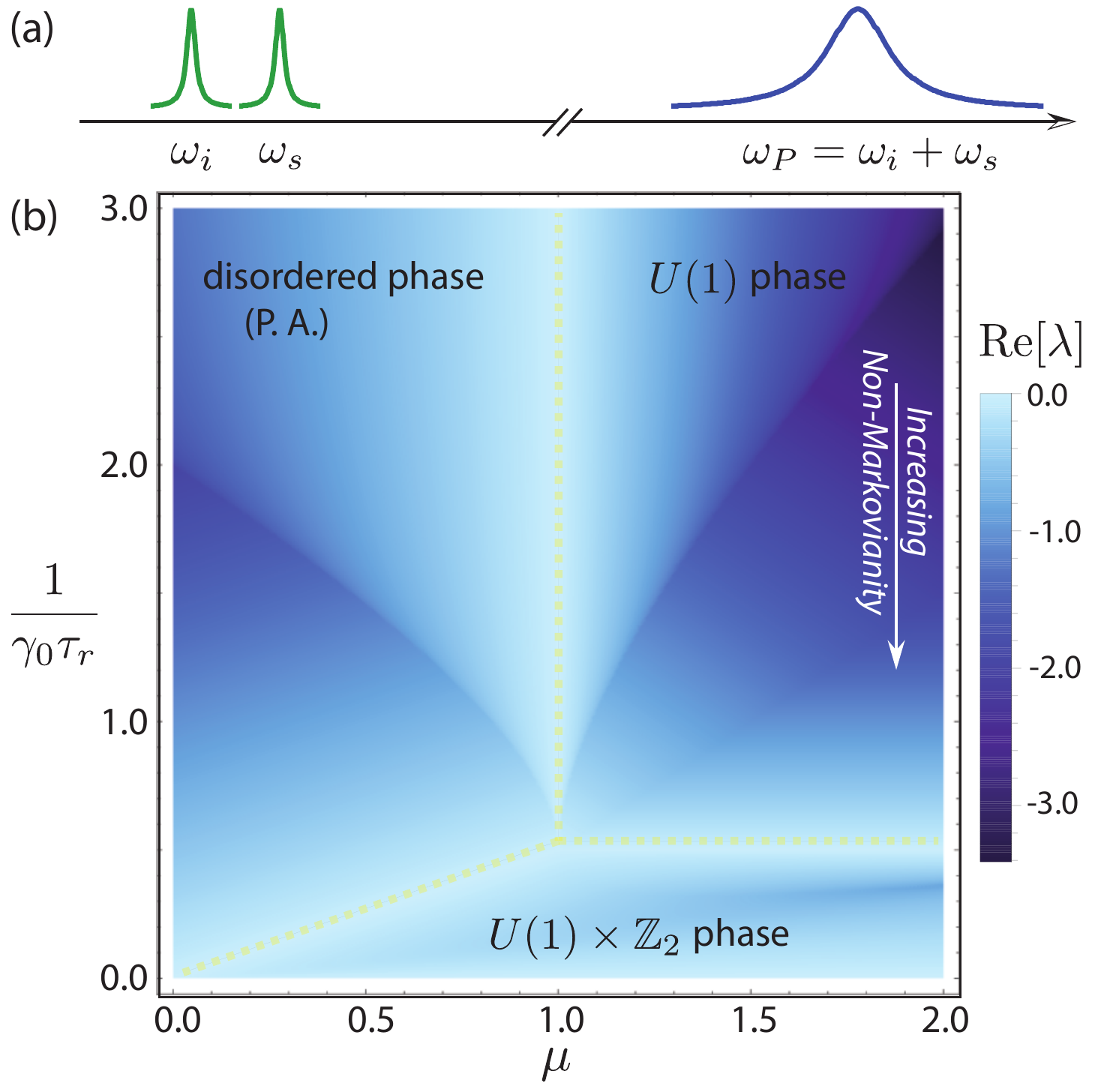}
\caption{(a) Schematic of the two-mode system. (b) The phase diagram as a function of the drive strength $\mu \equiv F_P/F_{cr}$ and the normalized
reservoir decay rate $(\gamma_0 \tau_r)^{-1}$. The color scale indicates the least negative real part of the eigenvalues of the susceptibility matrix (see text).
Critical points and phase boundaries (dashed lines) correspond to the vanishing of this real part, i.e. a divergent relaxation
time.  }
\label{fig:phasediagram}
\end{figure}
 
{\em Model.} The Hamiltonian of our system is given by \cite{patil2015,chakram2015}
\begin{equation}
\mathcal{H}/\hbar = \sum_{k} \omega_k \hat{a}^\dagger_k \hat{a}_k - \chi \hat{x}_P \hat{x}_i \hat{x}_s -  (F_P e^{-i \omega_P t} \hat{a}_P^\dagger + \mathrm{h.c.})  \nonumber
\end{equation}
where the indices $k = \{i,s,P\}$ denote the idler, signal and pump modes. The second term represents the two-mode interaction mediated by the actively driven pump, 
while the third term represents the classical drive at the pump frequency. The influence of the reservoir is incorporated through a master equation \cite{weiss} and leads to Heisenberg-Langevin equations of the form
\begin{eqnarray}
\dot{a}_i &=& -\frac{1}{2} \int_{-\infty}^t \gamma (t-t') a_i(t') dt' + i g a_s^\dagger a_P + i f_i \nonumber \\
\dot{a}_s &=& -\frac{1}{2} \int_{-\infty}^t \gamma (t-t') a_s(t') dt' + i g a_i^\dagger a_P + i f_s \nonumber \\
\dot{a}_P &=& -\frac{\gamma_P}{2} a_P + i g a_i a _s + i F_P \nonumber
\end{eqnarray}
where $\gamma(t)$ is the dissipation kernel in the rotating frame, and is related to the Langevin forces $f_{i,s}$ through the fluctuation-dissipation theorem, and the normalized coupling strength is $g = \chi x_{0,i} x_{0,s} x_{0,P}$ with $x_{0,\{i,s,P\}}$ denoting the zero point amplitudes of
the respective modes. In the above, we have made the rotating wave approximation, assumed that the pump mode is driven on resonance, i.e. $\omega_P = \omega_i + \omega_s$ and that the damping rate of the pump mode 
$\gamma_P$ is much larger than those of the signal and idler modes. 
For a Markovian reservoir, i.e. $\gamma(t) = \gamma_0 \delta(t)$, this system exhibits a continuous transition at a critical pump 
amplitude $F_{cr} = \frac{\gamma_P \gamma_0}{4 g}$ from a disordered (parametric amplifier) phase to an ordered phase characterized by parametric 
self-oscillation of the signal and idler modes and a spontaneous breaking of the $U(1)$ symmetry related to the 
difference between signal and idler phases \cite{drummond2005, chakram2015}. 

In this work, we consider the case where the signal and idler modes are in contact with a non-Markovian reservoir with a dissipation kernel $\gamma(t) = \gamma_0 \frac{e^{- t/\tau_r}}{\tau_r}$, where $\tau_r$ represents the coherence time or `memory' of the reservoir. This form of non-equilibrium noise arises naturally in the context of several cavity optomechanical systems \cite{wilsonrae2008, patil2015, chakram2015, groeblacher2015} as well as hybrid systems in which an optomechanical system is coupled to 
coherent ensembles of quantum spins \cite{bariani2014,bariani2015}. 

\begin{figure}[b]
\includegraphics[width=0.5\textwidth]{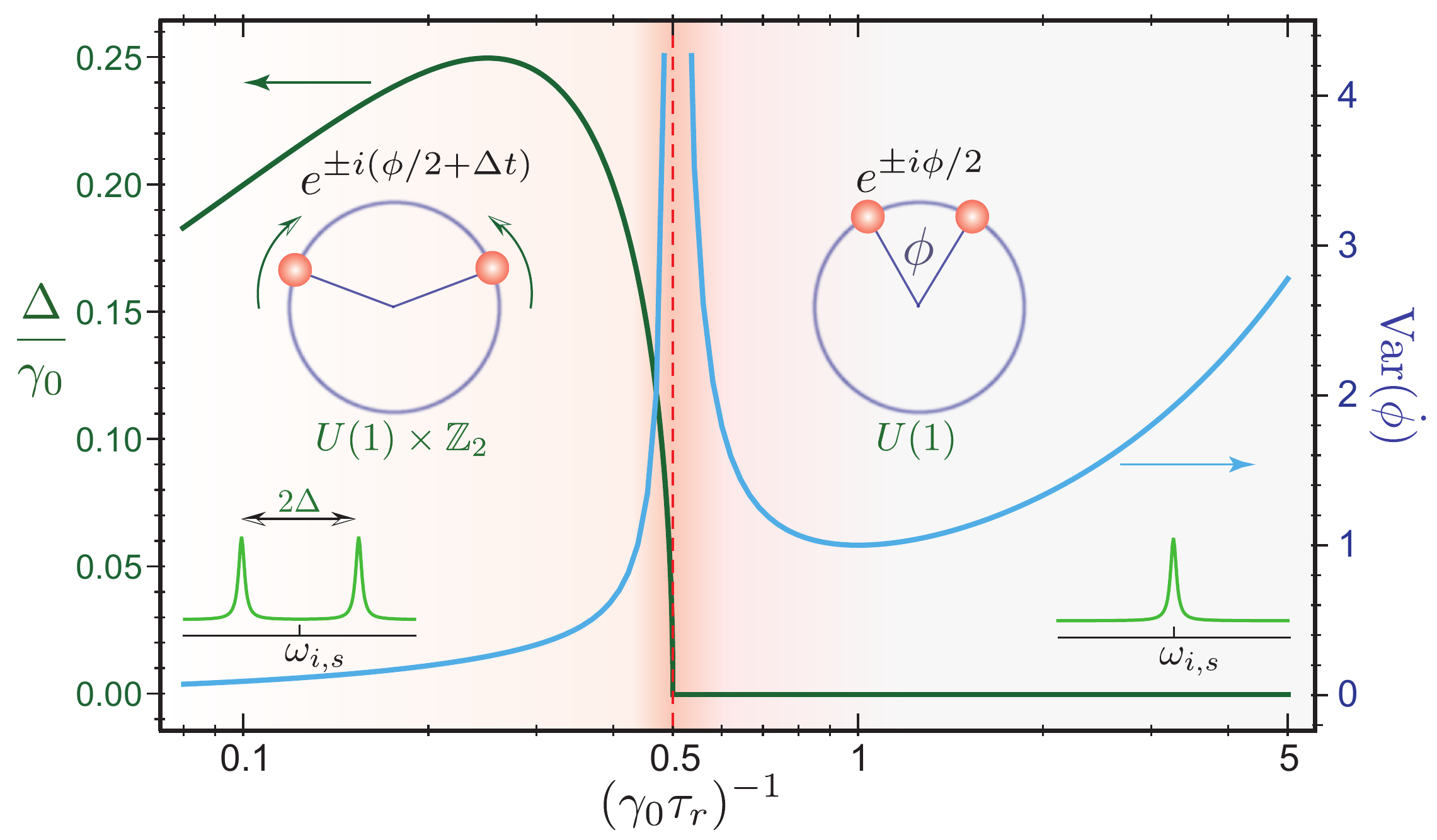}
\caption{The transition between the $U(1)$ and the $U(1) \times \mathbb{Z}_2$ phase versus the normalized reservoir decay rate, $(\gamma_0 \tau_r)^{-1}$. The critical point occurs at $(\gamma_0 \tau_r)^{-1} = \frac{1}{2}$, corresponding to a divergent variance, $\mathrm{Var}(\dot{\phi})$,  
of the instantaneous frequency of the signal and idler modes. Below this critical point, these modes no longer self-oscillate at their nominal resonances but shift to $\omega_{i} \rightarrow \omega_i \pm \Delta, \omega_{s} \rightarrow \omega_s \mp \Delta$, corresponding to a breaking of the $\mathbb{Z}_2$ symmetry (see inset, bottom). In contrast to the spontaneously chosen but constant phase difference between the two modes in the $U(1)$ phase, the phases of these modes now oscillate (see inset, top) at a frequency $\Delta$ that continuously grows from zero below the critical point.   }
\label{fig:phasediagram}
\end{figure}
\begin{figure*}[t]
\centering
\includegraphics[width=0.90\textwidth]{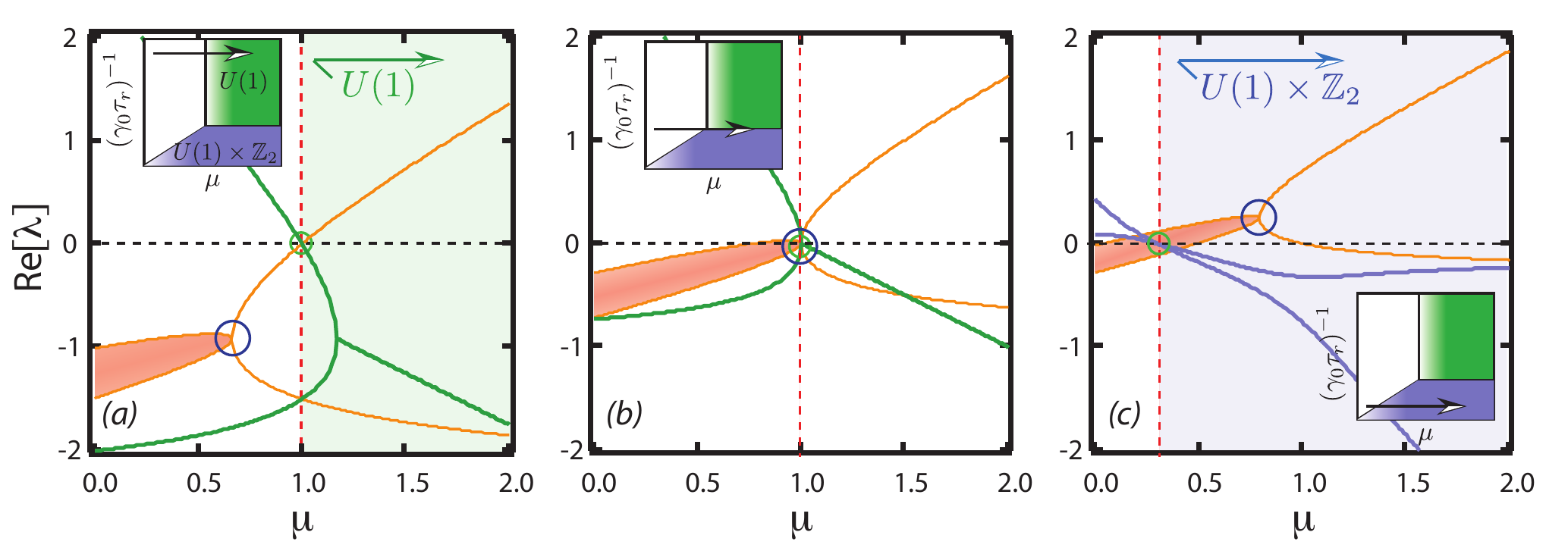}
\caption{ The behavior of the low lying eigenspectrum corresponding to the disordered phase (orange), $U(1)$ phase (green) and $U(1) \times \mathbb{Z}_2$ phase (blue) 
with increasing reservoir coherence time $\tau_r$ showing the relative
positions of the exceptional point (blue circle) and the critical point (green circle) {\em vs} the drive strength. The imaginary part of the eigenvalue 
is represented as the width of the eigenmode. The exceptional point corresponds to a coalescence of the eigenvalues and eigenmodes and a vanishing
imaginary part. The critical point occurs when the disordered phase becomes unstable ($\mathrm{Re}[\lambda] >0$) and gives way to the broken symmetry phases. 
(a) In the Markovian regime, i.e. $(\gamma_0 \tau_r)^{-1} \gg \frac{1}{2}$, the exceptional point occurs before the critical point
governing the transition to the $U(1)$ phase. The eigenvalues are purely real in the vicinity of the critical point. 
(b) At $(\gamma_0 \tau_r)^{-1} = \frac{1}{2}$, the exceptional point and the critical point coincide, i.e. the real
and imaginary parts of the eigenvalues vanish simultaneously at the critical point, indicating the emergence of the $U(1) \times \mathbb{Z}_2$ phase. 
(c) Deep in the non-Markovian regime, i.e. $(\gamma_0 \tau_r)^{-1} \ll \frac{1}{2}$, the critical point occurs before the exceptional point and the 
transition to the $U(1) \times \mathbb{Z}_2$ phase occurs when the eigenvalues are purely imaginary. The displayed eigenspectra correspond to (a) $(\gamma_0 \tau_r)^{-1} = 1.25$, (b) $(\gamma_0 \tau_r)^{-1} = 0.50$, and (c) $(\gamma_0 \tau_r)^{-1} = 0.15$. }
\label{fig:figschem}
\end{figure*}

{\em Mean field solutions and phase diagram.} The Heisenberg-Langevin equations can be cast in Fourier space as $- i \omega {\bf a} = {\bf \Sigma} {\bf a} + {\bf v}$ (see Supplemental Information). Here, ${\bf a} = (a_i, a_s, a_P)^T$ and the noise forces ${\bf v}$ are zero-mean gaussian variables whose correlation function is related to the dissipation kernel via the fluctuation-dissipation theorem. 
The eigenvalues $\lambda$ of the susceptibility matrix ${\bf \Sigma} + i \omega {\bf I}$ determine the low energy eigenspectrum and phase diagram of this system. Phase boundaries between dynamical states of distinct symmetries are associated with a vanishing of the least negative real part of eigenvalues of the susceptibility matrix, i.e. a divergent relaxation time \cite{kessler2012, honing2012}. Away from these phase boundaries, steady-state solutions for the signal and idler modes are represented by the form $a_{i,s} = |a_{i,s}| e^{i \theta_{i,s}} e^{-i \Delta_{i,s} t}$, and the stability of these solutions to generic perturbations is indicated by non-positive real parts of the eigenvalues (see Supplemental Information).
The mean field solutions of this non-Markovian system allow for three stable dynamical phases (Fig. 1). 

We first consider the regime of small reservoir coherence time, i.e. $(\gamma_0 \tau_r)^{-1} \gg 1$. In this regime, the eigenvalues of the disordered phase are $\lambda_\pm = \frac{\gamma_0}{4} \left[ (\mu - \frac{2}{\gamma_0 \tau_r}) \pm \sqrt{(\mu + \frac{2}{\gamma_0 \tau_r})^2 - \frac{8}{\gamma_0 \tau_r}}\right]$ where $\mu = F_P/F_{cr}$ is the normalized pump drive. The eigenvalues are purely real and the dynamical matrix exhibits near-Markovian behavior. For $\mu < 1$, the mean amplitudes of the signal and idler modes are zero. This corresponds to the disordered or parametric amplifier phase. For $\mu > 1$, the signal and idler modes exhibit parametric self-oscillation and the steady-state solutions are $a_{i,s} = i \frac{\sqrt{\gamma_0 \gamma_P}}{2 g}e^{\pm i \phi/2} \sqrt{\mu - 1}$. The $U(1)$ symmetry corresponding to the unconstrained phase difference $\phi$ between the signal and idler modes is spontaneously broken at the phase boundary $\mu_{cr} = 1$. As such, we denote this as the $U(1)$ phase. 
These solutions are consistent with the corresponding phases in a purely Markovian system. 

In contrast, as the coherence time $\tau_r$ is increased, the system is qualitatively modified due to the competing effects of intrinsic damping and reservoir coherence. As the timescales of these processes become commensurate, the eigenvalues morph into complex conjugate pairs analogous to $\mathcal{PT}$ symmetry breaking of the dynamical matrix. For $(\gamma_0 \tau_r)^{-1} < \frac{1}{2}$, a new steady-state solution emerges given by $a_{i,s} \propto i e^{\pm i (\phi/2 + \Delta t)} \sqrt{\mu - \mu_{cr}}$ with
$\Delta = \tau_r^{-1} \sqrt{\frac{\gamma_0 \tau_r}{2} - 1}$. Further, the critical drive strength monotonically decreases as $\mu_{cr} = \frac{2}{\gamma_0 \tau_r}$. 
In this new phase, the signal and idler modes exhibit self-oscillatory behavior not at their nominal resonances but at shifted frequencies $\omega_{i} \rightarrow \omega_{i} \pm \Delta, \omega_{s} \rightarrow \omega_{s} \mp \Delta$, with the choice of $\pm \Delta$ corresponding to a spontaneous breaking of a $\mathbb{Z}_2$ symmetry. In contrast to the fixed phase difference between the signal and idler modes in the $U(1)$ phase, the phases of these modes now oscillate at a rate $\Delta$. To further establish that this is a distinct phase, we calculate the dynamical states for a fixed drive $\mu > 1$ as the reservoir coherence time is reduced (Fig. 2). We find that the reservoir-induced frequency shift $\Delta$ continuously grows from zero below the phase boundary $(\gamma_0 \tau_r)^{-1} = \frac{1}{2}$, while the variance of the difference-phase fluctuations, $\mathrm{Var}(\dot{\phi})$, diverges at the phase boundary, characteristic of a continuous phase transition with an order parameter $\Delta$. 

The emergence of the novel $U(1) \times \mathbb{Z}_2$ phase is intimately related to the appearance of exceptional points of the dynamical matrix. At these points, the eigenvalues and eigenmodes of the dynamical matrix coalesce, signifying the transition from purely real eigenvalues to complex eigenvalues, resulting in distinct topological properties akin to a Berry phase in the vicinity of such points \cite{kato1966,Heiss2012}. At the phase boundary $(\gamma_0 \tau_r)^{-1} = \frac{1}{2}$, the critical point $\mu_c=1$ coincides with the exceptional point and the real and imaginary parts of $\lambda_{\pm}$ simultaneously vanish (Fig. 3), leading to non-reciprocal behavior and the simultaneous breaking of a discrete ($\mathbb{Z}_2$) symmetry in addition to the $U(1)$ symmetry related to the signal-idler phase difference. 

\begin{figure}[t]
\includegraphics[width=0.35\textwidth]{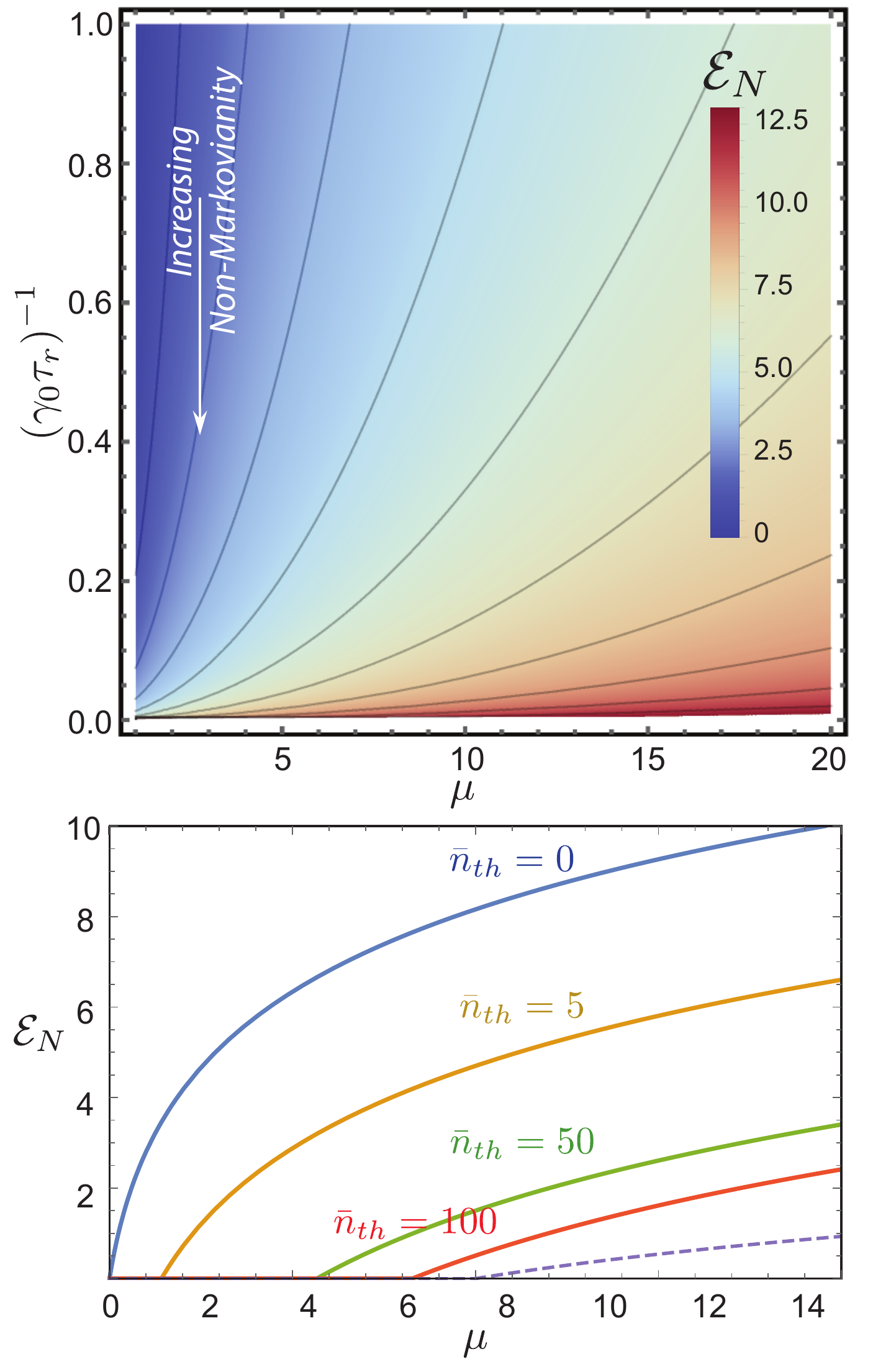}
\caption{(Top) The logarithmic negativity $\mathcal{E}_N$ as a measure of the bipartite entanglement \cite{peres1996, plenio2005} between the signal and idler modes 
{\em vs} the drive strength $\mu$ and the normalized reservoir decay rate $(\gamma_0 \tau_r)^{-1}$. (Bottom) In the $U(1) \times \mathbb{Z}_2$ phase, 
the entanglement between the two modes extends well beyond the quantum regime and can be observed even for large thermal occupancy of the
two modes. The logarithmic negativity is shown for increasing thermal occupancy $\bar{n}_{th}$ {\em vs} drive strength for $(\gamma_0 \tau_r)^{-1}= \frac{1}{5}$.
For comparison, the logarithmic negativity for a Markovian system with $\bar{n}_{th} = 5$ is shown in dashed lines.}
\label{fig:lognegativity}
\end{figure}

The appearance of the exceptional point near the phase boundary is accompanied by an enhanced degree of
entanglement between the two modes in the vicinity of the phase boundary between the $U(1)$ and $U(1) \times \mathbb{Z}_2$ phase. 
Interestingly, the squeezed variance of these cross-quadratures scales as $(\bar{n}_{th} + \frac{1}{2}) \frac{2 (\gamma_0 \tau_r)^{-1}}{(1 + \mu)(2 (\gamma_0 \tau_r)^{-1} + \mu)} \propto \frac{1}{\mu^2}$ for large drive strength, in contrast to the Markovian scaling $(\bar{n}_{th} + \frac{1}{2}) \frac{1}{1 + \mu}\propto \frac{1}{\mu}$, where $\bar{n}_{th}$ is the average thermal population of the signal and idler modes (see Supplemental Information, \cite{chakram2015}, \cite{cheung2016}). This enhancement over a Markovian system is also reflected in the logarithmic negativity. As can be seen in Fig. 4, this enhanced degree of entanglement persists even at large thermal occupancy of the signal and idler modes. We speculate that this enhancement is due to the topological properties of the exceptional point and the non-reciprocal behavior of the system in its vicinity. 

{\em Conclusions. } In summary, we identify a class of non-Markovianity that results in the
emergence of a novel broken symmetry phase in a driven, dissipative quantum system. We analyze the phase diagram of
this open quantum system and show that the emergent phase is accompanied by the appearance of exceptional
points in the system. This emergent phase manifests a 
larger degree of two-mode entanglement than would be observed in a Markovian system. We note that the two-mode 
system and the form of non-Markovianity considered here are readily accessible in cavity optomechanical systems as well
as various hybrid quantum systems, paving the way for experimental demonstrations of these predictions well into the quantum regime. 
Future work will extend this analysis to the regime of spatially multimode optomechanical systems and discuss the interplay 
between non-Markovian correlations, optomechanical synchronization, spatial fluctuations and driven, dissipative dynamics. 
In addition to realizing metrologically relevant optomechanical states, we suggest that 
this interplay also offers a new arena for disorder-free optomechanical realizations of dynamical phases with novel 
broken symmetries such as have been recently observed in spin systems \cite{choi2017, zhang2017} . 

We gratefully acknowledge A. Chandran, A. Polkovnikov and T. Villazon for valuable discussions and comments on the manuscript. 
This work was supported by the DARPA QuASAR program through a grant from the ARO, the ARO MURI on non-equilibrium 
Many-body Dynamics (63834-PH-MUR) and an NSF INSPIRE award. M. V. acknowledges support from the Alfred P. Sloan 
Foundation.

\bibliographystyle{unsrt}
\bibliography{OpenQuantumBib}

\begin{thebibliography}{10}

\bibitem{rivas2011}
{\'A}.~Rivas and S.~F. Huelga.
\newblock Open quantum systems : An introduction.
\newblock {\em arXiv:1104.5242}, 2011.

\bibitem{schindler2012}
P.~Schindler, M.~M\"{u}ller, D.~Nigg, J.~T. Barreiro, E.~A. Martinez,
  M.~Hennrich, T.~Monz, S.~Diehl, P.~Zoller, and R.~Blatt.
\newblock Quantum simulation of open-system dynamical maps with trapped ions.
\newblock {\em Nature Phys.}, 9:361, 2012.

\bibitem{xiang2013}
Z.~Xiang, S.~Ashhab, J.~Q. You, and F.~Nori.
\newblock Hybrid quantum circuits : Superconducting circuits interacting with
  other quantum systems.
\newblock {\em Rev. Mod. Phys.}, 85:623, 2013.

\bibitem{aspelmeyer2014}
M.~Aspelmeyer, T.~J. Kippenberg, and F.~Marquardt.
\newblock Cavity optomechanics.
\newblock {\em Rev. Mod. Phys.}, 86:1391, 2014.

\bibitem{kurizki2015}
G.~Kurizki, P.~Bertet, Y.~Kubo, K.~M{\o}lmer, D.~Petrosyan, P.~Rabl, and
  J.~Schmiedmayer.
\newblock Quantum technologies with hybrid systems.
\newblock {\em PNAS}, 112:3866, 2015.

\bibitem{diehl2008}
S.~Diehl, A.~Micheli, A.~Kantian, B.~Kraus, H.~P. B\"{u}chler, and P.~Zoller.
\newblock Quantum states and phases in driven open quantum systems with cold
  atoms.
\newblock {\em Nature Phys.}, 4:878, 2008.

\bibitem{muller2012}
M.~M\"{u}ller, S.~Diehl, G.~Pupillo, and P.~Zoller.
\newblock Engineered open systems and quantum simulations with atoms and ions.
\newblock {\em Adv. At. Mol. Opt. Phys.}, 61:1, 2012.

\bibitem{tomadin2012}
A.~Tomadin, S.~Diehl, M.~D. Lukin, P.~Rabl, and P.~Zoller.
\newblock Reservoir engineering and dynamical phase transitions in
  optomechanical arrays.
\newblock {\em Phys. Rev. A}, 86:033821, 2012.

\bibitem{drummond2005}
P.~D. Drummond and K.~Dechoum.
\newblock Universality of quantum critical dynamics in a planar optical
  parametric oscillator.
\newblock {\em Phys. Rev. Lett.}, 95:083601, 2005.

\bibitem{wouters2007}
M.~Wouters and I.~Carusotto.
\newblock Goldstone mode of optical parametric oscillators in planar
  semiconductor microcavities in the strong-coupling regime.
\newblock {\em Phys. Rev. A}, 76:043807, 2007.

\bibitem{emary2003}
C.~Emary and T.~Brandes.
\newblock Chaos and the quantum phase transition in the {D}icke model.
\newblock {\em Phys. Rev. E}, 67:066203, 2003.

\bibitem{baumann2010}
K.~Baumann, C.~Guerlin, F.~Brennecke, and T.~Esslinger.
\newblock Dicke quantum phase transition with a superfluid gas in an optical
  cavity.
\newblock {\em Nature}, 464:1301, 2010.

\bibitem{klinder2015}
J.~Klinder, H.~Kessler, M.~Wolke, L.~Mathey, and A.~Hemmerich.
\newblock Dynamical phase transition in the open {D}icke model.
\newblock {\em PNAS}, 112:3290, 2015.

\bibitem{dicke1954}
R.~H. Dicke.
\newblock Coherence in spontaneous radiation processes.
\newblock {\em Phys. Rev.}, 93:99, 1954.

\bibitem{lipkin1965}
H.~J. Lipkin, N.~Meshkov, and A.~J. Glick.
\newblock Validity of many-body approximation methods for a solvable model :
  (i). exact solutions and perturbation theory.
\newblock {\em Nucl. Phys.}, 62:188, 1965.

\bibitem{dechoum2004}
K.~Dechoum, P.~D. Drummond, S.~Chaturvedi, and M.~D. Reid.
\newblock Critical fluctuations and entanglement in the nondegenerate
  parametric oscillator.
\newblock {\em Phys. Rev. A}, 70:053807, 2004.

\bibitem{dallatorre2010}
E.~G. Dalla~Torre, E.~Demler, T.~Giamarchi, and E.~Altman.
\newblock Quantum critical states and phase transitions in the presence of
  non-equilibrium noise.
\newblock {\em Nature Phys.}, 6:806, 2010.

\bibitem{nagy2011}
D.~Nagy, G.~Szirmai, and P.~Domokos.
\newblock Critical exponent of a quantum-noise-driven phase transition : The
  open-system {D}icke model.
\newblock {\em Phys. Rev. A}, 84:043637, 2011.

\bibitem{nagy2016}
D.~Nagy and P.~Domokos.
\newblock Critical exponent of quantum phase transitions driven by colored
  noise.
\newblock {\em Phys. Rev. A}, 94:063862, 2016.

\bibitem{patil2015}
Y.~S. Patil, S.~Chakram, L.~Chang, and M.~Vengalattore.
\newblock Thermomechanical two-mode squeezing in an ultrahigh ${Q}$ membrane
  resonator.
\newblock {\em Phys. Rev. Lett.}, 115:017202, 2015.

\bibitem{chakram2015}
S.~Chakram, Y.~S. Patil, and M.~Vengalattore.
\newblock Multimode phononic correlations in a nondegenerate parametric
  amplifier.
\newblock {\em New Journal of Phys.}, 17:063018, 2015.

\bibitem{weiss}
U.~Weiss.
\newblock {\em Quantum dissipative systems}.
\newblock World Scientific, 2008.

\bibitem{wilsonrae2008}
I.~Wilson-Rae.
\newblock Intrinsic dissipation in nanomechanical resonators due to phonon
  tunneling.
\newblock {\em Phys. Rev. B}, 77:245418, 2008.

\bibitem{groeblacher2015}
S.~Groeblacher, A.~Trubarov, N.~Prigge, G.~D. Cole, M.~Aspelmeyer, and
  J.~Eisert.
\newblock Observation of non-{M}arkovian micro-mechanical {B}rownian motion.
\newblock {\em Nature Comm.}, 6:7606, 2015.

\bibitem{bariani2014}
F.~Bariani, S.~Singh, L.~F. Buchmann, M.~Vengalattore, and P.~Meystre.
\newblock Hybrid optomechanical cooling by atomic lambda systems.
\newblock {\em Phys. Rev. A}, 90:033838, 2014.

\bibitem{bariani2015}
F.~Bariani, H.~Soek, S.~Singh, M.~Vengalattore, and P.~Meystre.
\newblock Atom-based coherent quantum-noise cancellation in optomechanics.
\newblock {\em Phys. Rev. A}, 92:043817, 2015.

\bibitem{kessler2012}
E.~M. Kessler, G.~Giedke, A.~Imamoglu, S.~F. Yelin, M.~D. Lukin, and J.~I.
  Cirac.
\newblock Dissipative phase transition in a central spin system.
\newblock {\em Phys. Rev. A}, 86:012116, 2012.

\bibitem{honing2012}
M.~H\"{o}ning, M.~Moos, and M.~Fleischhauer.
\newblock Critical exponents of steady-state phase transitions in fermionic
  lattice models.
\newblock {\em Phys. Rev. A}, 86:013606, 2012.

\bibitem{kato1966}
T.~Kato.
\newblock {\em Perturbation theory for linear operators}.
\newblock Springer Berlin, 1966.

\bibitem{Heiss2012}
W.~D. Heiss.
\newblock The physics of exceptional points.
\newblock {\em J. Phys. A: Math. Theor.}, 45:444016, 2012.

\bibitem{peres1996}
A.~Peres.
\newblock Separability criterion for density matrices.
\newblock {\em Phys. Rev. Lett.}, 77:1413, 1996.

\bibitem{plenio2005}
M.~B. Plenio.
\newblock Logarithmic {N}egativity : A full entanglement monotone that is not
  convex.
\newblock {\em Phys. Rev. Lett.}, 95:090503, 2005.

\bibitem{cheung2016}
H.~F.~H. Cheung, Y.~S. Patil, L.~Chang, S.~Chakram, and M.~Vengalattore.
\newblock Nonlinear phonon interferometry at the {H}eisenberg limit.
\newblock {\em arXiv:1601.02324}, 2016.

\bibitem{choi2017}
S.~Choi, J.~Choi, R.~Landig, G.~Kucsko, H.~Zhou, J.~Isoya, F.~Jelezko,
  S.~Onoda, H.~Sumiya, V.~Khemani, C.~von Keyserlingk, N.~Y. Yao, E.~Demler,
  and M.~D. Lukin.
\newblock Observation of discrete time-crystalline order in a disordered
  dipolar many-body system.
\newblock {\em Nature}, 543:221, 2017.

\bibitem{zhang2017}
J.~Zhang, P.~W. Hess, A.~Kyprianides, P.~Becker, A.~Lee, J.~Smith, G.~Pagano,
  I.-D. Potirniche, A.~C. Potter, A.~Vishwanath, N.~Y. Yao, and C.~Monroe.
\newblock Observation of a discrete time crystal.
\newblock {\em Nature}, 543:217, 2017.

\end{thebibliography}

\newpage
\onecolumngrid

\section*{Supplemental Information}
The Hamiltonian for the two-mode driven dissipative system is given by \cite{drummond2005, patil2015, chakram2015}
\begin{equation}
\mathcal{H}/\hbar = \sum_{k = \{i,s,P\}} \omega_k \hat{a}_k^\dagger \hat{a}_k - \chi \hat{x}_P \hat{x}_i \hat{x}_s - (F_P e^{-i \omega_P t} \hat{a}_P^\dagger + \mathrm{h.c.})
\end{equation}
In the interaction picture with $\mathcal{H}_0/\hbar = \sum_k \omega_k \hat{a}_k^\dagger \hat{a}_k$, and making the rotating wave approximation, the interaction Hamiltonian transforms to $\mathcal{H}/\hbar = - g (\hat{a}_s^\dagger \hat{a}_i^\dagger \hat{a}_P + \hat{a}_P^\dagger \hat{a}_s \hat{a}_i) - (F_P \hat{a}_P^\dagger + \mathrm{h.c.})$ where $g = \chi x_{0,i} x_{0,s} x_{0,P}$ and $x_{0,k}$ denotes the zero-point amplitude of the respective modes. Here, we have assumed that the pump mode is actuated by a resonant, classical force and that $\omega_P = \omega_i + \omega_s$. 

Further, the influence of the reservoir on these modes is incorporated through noise operators $\hat{f}$ and takes the form $\mathcal{H}_r/\hbar = -\sum_k (\hat{a}_k^\dagger \hat{f}_k + \mathrm{h.c.})$. For the signal and idler modes, these noise forces are zero-mean, gaussian random variables whose two-point correlation is related to the dissipation kernel $\gamma(t)$ in accordance with the fluctuation dissipation theorem. Here, we assume that the signal and idler modes are in contact with a colored reservoir with a dissipation kernel given by $\gamma(t - t') = \gamma_0 \tau_r^{-1} \exp(-(t-t')/\tau_r) \Theta(t-t')$ where $\Theta(t)$ is the Heaviside step function. Accordingly, these noise forces satisfy the following relations, $\langle f_k \rangle = 0$, and $\langle f_k(t) f_l^\dagger(t') \rangle = \delta_{kl} \times (\bar{n}_{th} + 1) \frac{\gamma_0}{2 \tau_r} e^{-|t-t'|/\tau_r}$ where $\bar{n}_{th} = (\exp(\frac{\hbar \omega_{i,s}}{k_B T}) - 1)^{-1}$. 
 
In accordance with typical experimental situations in optomechanical systems \cite{patil2015}, we assume that the pump mode is in contact with a Markovian reservoir and that its damping rate is much larger than those of the signal and idler modes, i.e. $\gamma_P \gg \gamma_0$. This leads to the Heisenberg-Langevin equations of the form
\begin{eqnarray}
\dot{a}_i &=& -\frac{1}{2}\int_{-\infty}^t \gamma(t-t') a_i(t') dt' + i g a_s^\dagger a_P + i f_i \\
\dot{a}_s &=& -\frac{1}{2}\int_{-\infty}^t \gamma(t-t') a_s(t') dt' + i g a_i^\dagger a_P + i f_s \\
\dot{a}_P &=& -\frac{\gamma_P}{2} a_P + i g a_i a_s + i F_P
\end{eqnarray}
Here, we are ignoring the Langevin forces on the pump. As explained in \cite{chakram2015}, the pump noise can be ignored in evaluating the dynamical steady-state phases or the 
degree of two-mode correlations below threshold. Above threshold, this pump noise has an appreciable effect on the two-mode squeezing. In our calculations of the squeezing spectra
above threshold, this pump noise is included by assuming that the pump mode is in contact with a Markovian reservoir as explained in \cite{chakram2015}. 

These equations can be recast by defining the dimensionless amplitudes $A_{i,s} = a_{i,s} \frac{2 g}{\sqrt{\gamma_0 \gamma_P}}$ and $A_P = a_P \frac{2 g}{\gamma_0}$ to obtain
\begin{eqnarray}
\dot{A}_i &=& \frac{1}{2}\left[ -\int_{-\infty}^t \gamma(t-t') A_i(t') dt' + i \gamma_0 A^*_s A_P + i \gamma_0 \tilde{f}_i \right]\\
\dot{A}_s &=& \frac{1}{2}\left[ -\int_{-\infty}^t \gamma(t-t') A_s(t') dt' + i \gamma_0 A^*_i A_P + i \gamma_0 \tilde{f}_s \right]\\
\dot{A}_P &=& \frac{1}{2}\left[ -\gamma_P A_P(t) + i \gamma_P A_i A_s + i \gamma_P \mu \right]
\end{eqnarray}
where $\tilde{f}_{i,s} = \gamma_0^{-1} \frac{4 g}{\sqrt{\gamma_0 \gamma_P}} f_{i,s}$ and we have defined the normalized drive strength $\mu = F_P/F_{cr}$ where $F_{cr} = \frac{\gamma_P \gamma_0}{4 g}$.
\subsection*{Steady state dynamical phases and mean field phase diagram}
We consider the situation where the signal and idler modes are not driven, i.e. they are only subject to the Langevin forces originating from their coupling to the colored reservoir. In contrast, the pump mode is actively driven by a classical force represented by the normalized drive $\mu$. In various regimes of the drive strength $\mu$ and the reservoir coherence time $\tau_r$, we consider dynamical steady-state phases represented by the ansatz 
\begin{eqnarray}
A_{i,s} &=& \bar{A}_{i,s} e^{-i \Delta_{i,s} t} \\
A_P &=& \bar{A}_P
\end{eqnarray}
Substituting this ansatz into Eqs.(5-7), we obtain
\begin{eqnarray}
-i \Delta_i \bar{A}_i e^{-i \Delta_i t} &=& -\frac{1}{2} e^{-i \Delta_i t} \bar{A}_i \tilde{\gamma}(\Delta_i) + i \frac{\gamma_0}{2} \bar{A}^*_s \bar{A}_P e^{i \Delta_s t} \\
-i \Delta_s \bar{A}_s e^{-i \Delta_s t} &=& -\frac{1}{2} e^{-i \Delta_s t} \bar{A}_s \tilde{\gamma}(\Delta_s) + i \frac{\gamma_0}{2} \bar{A}^*_i \bar{A}_P e^{i \Delta_i t} \\
0 &=& -\frac{1}{2} \gamma_P \bar{A}_P + i \frac{\gamma_P}{2} \bar{A}_i \bar{A}_s e^{-i (\Delta_i + \Delta_s) t} + i \frac{\gamma_P}{2} \mu
\end{eqnarray}
Here, we have used the fourier transform of the dissipation kernel, $\tilde{\gamma}(\omega) = \int dt \gamma(t) e^{i \omega t} = \gamma_0 (1 - i \omega \tau_r)^{-1}$. 

These equations always admit the trivial solution $\bar{A}_i = \bar{A}_s = 0, \bar{A}_P = i \mu$. For dynamical steady-states with finite signal and idler amplitudes, the above equations require $\Delta_i + \Delta_s = 0$. Hence, below we define $\Delta \equiv \Delta_i = -\Delta_s$. 
Eqs.(10-11) together yield the following condition
\begin{equation}
\left( \frac{\tilde{\gamma}(\Delta)}{2} - i \Delta \right) \left( \frac{\tilde{\gamma}(-\Delta)}{2} + i \Delta \right)^* \bar{A}_i = \frac{\gamma_0^2}{4} |\bar{A}_P|^2 \bar{A}_i
\end{equation}
Since $\tilde{\gamma}(-\omega) = \tilde{\gamma}^*(\omega)$, this requires steady-state phases with non-zero signal and idler mode amplitudes to satisfy the condition
\begin{equation}
\left( \frac{\tilde{\gamma}(\Delta)}{2} - i \Delta \right)^2 = \frac{\gamma_0^2}{4} |\bar{A}_P|^2
\end{equation}
indicating that $\frac{\tilde{\gamma}_0}{2} \equiv \frac{\tilde{\gamma}(\Delta)}{2} - i \Delta$ is real and positive. Further, 
Eqs.(10,11) in combination with Eq(12) yields the following expression for the signal and idler amplitudes,
\begin{eqnarray}
\left( \frac{\tilde{\gamma}_0}{\gamma_0} \right)^2 \left[ 1 + 2 \left| \frac{\gamma_0}{\tilde{\gamma}_0}\right| |\bar{A}_{i,s}|^2 + \left| \frac{\gamma_0}{\tilde{\gamma}_0} \right|^2 |\bar{A}_{i,s}|^4 \right] &=& \mu^2 \\
\Rightarrow |\bar{A}_{i,s}| &=& \sqrt{\mu - \frac{\tilde{\gamma}_0}{\gamma_0}} 
\end{eqnarray}
Accordingly, we define the critical pump amplitude $\mu_{cr} = \tilde{\gamma}_0/\gamma_0$ as the drive strength beyond which the signal and idler modes develop a non-zero amplitude, i.e. the onset of parametric self-oscillation. 

Lastly, given the constraint from Eq(13) that $\tilde{\gamma}_0/2$ be real-valued and positive, we obtain
\begin{equation}
\frac{1}{2} \tilde{\gamma}(\Delta) - i \Delta = \frac{1}{2} \frac{\gamma_0}{1 + \tau_r^2 \Delta^2} + i \Delta \left( \frac{1}{2} \frac{\gamma_0 \tau_r}{1 + \tau_r^2 \Delta^2} - 1 \right) \in \mathbb{R}
\end{equation}
yielding $\Delta = 0$, or $\Delta = \tau_r^{-1} \sqrt{\frac{\gamma_0 \tau_r}{2} - 1}$. Note that the latter solution is only meaningful for $\gamma_0 \geq 2 \tau_r^{-1}$.

Based on these relations, we can identify three distinct dynamical phases in this system. 
\begin{itemize}
\item In the regime $\gamma_0 \leq 2 \tau_r^{-1}$, the coherence time of the reservoir is small compared to the intrinsic damping time of the signal/idler modes. Here, we obtain the condition $\Delta = 0$ and $\tilde{\gamma}_0 = \tilde{\gamma}(0) = \gamma_0$. Hence, the critical drive strength is given by $\mu_{cr} = 1$. In this regime, for drive strengths $\mu < 1$, the only stable phase is the trivial solution $\bar{A}_i = \bar{A}_s = 0, \bar{A}_P = i \mu$. This is the disordered or parametric amplifier phase. As the drive strength is increased beyond $\mu_{cr} = 1$, the parametric amplifier phase becomes unstable (Fig. 3(a)) and gives way to the parametric oscillator phase characterized by $\bar{A}_{i,s} = i e^{\pm i \phi/2} \sqrt{\mu - 1}, \bar{A}_P = i$. The signal-idler phase difference $\phi$ is unconstrained and the emergence of this parametric oscillator phase is accompanied by the spontaneous breaking of the $U(1)$ symmetry associated with the choice of this phase. As such, we denote this to be the $U(1)$ phase. 

\item In the regime $\gamma_0 > 2 \tau_r^{-1}$, the coherence time of the reservoir is long compared to the intrinsic damping time of the signal/idler modes. As seen from Eqs(16,17), in this regime the critical point shifts to $\mu_{cr} = 2 (\gamma_0 \tau_r)^{-1} < 1$. For $\mu < \mu_{cr}$, the only stable phase is the disordered or trivial solution with $\bar{A}_{i,s} = 0$. For drive strengths $\mu_{cr} < \mu < 1$, the disordered phase is unstable and gives way to a self-oscillating phase with non-zero $\Delta$, given by $\bar{A}_{i,s} = i e^{\pm i (\phi/2 + \Delta t)} \sqrt{\mu - \mu_{cr}}, \bar{A}_P = i \mu_{cr}$ with $\Delta = \tau_r^{-1} \sqrt{\frac{\gamma_0 \tau_r}{2} - 1}$. In this dynamical phase, the signal and idler modes undergo self-oscillation at frequencies that are shifted away from their nominal frequencies by an amount $\Delta$. In addition to the breaking of the $U(1)$ symmetry associated with the choice of the signal-idler phase difference $\phi$, this phase also breaks the discrete $\mathbb{Z}_2$ symmetry associated with the sign of the frequency shift $\Delta$. As such, we denote this dynamical phase as the $U(1) \times \mathbb{Z}_2$ phase. For $\mu > 1$, all three solutions exist but the trivial solution and the $U(1)$ solution are unstable, with the $U(1) \times \mathbb{Z}_2$ solution remaining as the only stable dynamical phase. 
\end{itemize}

These three dynamical phases along with the phase boundaries demarcating these phases are shown in Fig.(1) of the main text. 
\subsection*{Exceptional points and stability of mean field dynamical phases}

The stability of the mean-field dynamical phases to generic perturbations is demonstrated by evaluating the eigenvalues of the susceptibility matrix $\bf{\Sigma} + i \omega {\bf I}$ as discussed in the main text. In particular, a stable dynamical phase is indicated by a susceptibility matrix whose eigenvalues have non-positive real parts. We outline the calculation of these eigenvalues for each dynamical phase below. We first distinguish between the mean amplitudes and the fluctuations by writing $A_{i,s} = (\bar{A}_{i,s} + \delta A_{i,s})$ with the mean amplitudes in each dynamical phase given by the expressions in the previous section. 
The equations of motion Eqs(5-7) yield
\begin{eqnarray}
\partial_t \left( \begin{array}{c} \delta A_i \\ \delta A_s \\ \delta A_P\end{array}\right) &=& \int_{-\infty}^t dt' \left( \begin{array}{ccc} -\frac{1}{2}\gamma(t-t') & 0 & i \frac{\gamma_0}{2} \bar{A}_s^* \delta(t-t') \\ 0 & -\frac{1}{2} \gamma(t-t') & i \frac{\gamma_0}{2} \bar{A}_i^* \delta(t-t') \\ i \frac{\gamma_P}{2} \bar{A}_s \delta(t-t') & i \frac{\gamma_P}{2} \bar{A}_i \delta(t-t') & -\frac{\gamma_P}{2} \delta(t-t') \end{array}\right) \left( \begin{array}{c} \delta A_i (t') \\ \delta A_s (t') \\ \delta A_P (t') \end{array}\right)\nonumber \\
 &+& \left( \begin{array}{ccc} 0 & i \frac{\gamma_0}{2} \bar{A}_P & 0 \\ i \frac{\gamma_0}{2} \bar{A}_P & 0 & 0 \\ 0 & 0 & 0 \end{array}\right) \left( \begin{array}{c} \delta A_i^* \\ \delta A_s^* \\ \delta A_P^* \end{array} \right) + \frac{i}{2} \left( \begin{array}{c} \gamma_0 \tilde{f}_i(t) \\ \gamma_0 \tilde{f}_s(t) \\ \gamma_P \mu \end{array} \right)
\end{eqnarray}
As shown in \cite{chakram2015}, the complex fluctuations can be decomposed into real quadratures in the form $\delta {\bf A} = \delta \vec{\alpha} + i \delta \vec{\beta}$ such that the above equation can be recast as 
\begin{eqnarray}
\delta \dot{\vec{\alpha}} &=& \int_{-\infty}^t dt' {\bf M}_\alpha(t-t') \delta \vec{\alpha}(t') + {\bf v}_\alpha(t) \\
\delta \dot{\vec{\beta}} &=& \int_{-\infty}^t dt' {\bf M}_\beta(t-t') \delta \vec{\beta}(t') + {\bf v}_\beta(t)
\end{eqnarray}
where ${\bf v}_{\alpha, \beta}$ are the Langevin noise terms and 
\begin{equation}
{\bf M}_{\alpha, \beta}(t) = \frac{1}{2} \left( \begin{array}{ccc} -\gamma(t) & \mp \gamma_0 |\bar{A}_P| \delta(t) & \gamma_0 |\bar{A}_s| \delta(t) \\ \mp \gamma_0 |\bar{A}_P| \delta(t) & -\gamma(t) & \gamma_0 |\bar{A}_i| \delta(t) \\ -\gamma_P |\bar{A}_s| \delta(t) & -\gamma_P |\bar{A}_i| \delta(t) & -\gamma_P \delta(t) \end{array} \right)
\end{equation}
Further, we define cross-quadratures of the signal and idler modes according to the relations $x_{\pm} = (\alpha_i \pm \alpha_s)/\sqrt{2}, y_\pm = (\beta_i \pm \beta_s)/\sqrt{2}$ such that the two-mode correlations due to parametric down-conversion are manifest as amplification and squeezing of the above quadratures. The fluctuations of these cross-quadratures are related to the original quadrature fluctuations $\delta \vec{\alpha}, \delta \vec{\beta}$ via the relations
\begin{equation}
\delta {\bf X} = {\bf R} \delta \vec{\alpha}; \,\,\, \delta {\bf Y} = {\bf R} \delta \vec{\beta}; \,\,\, {\bf R} = \frac{1}{\sqrt{2}} \left( \begin{array}{ccc} 1 & 1 & 0 \\ 1 & -1 & 0 \\ 0 & 0 & \sqrt{2} \end{array}\right)
\end{equation}
where $\delta {\bf X} = (\delta x_+, \delta x_-, \delta x_P)^T, \delta {\bf Y} = (\delta y_+, \delta y_-, \delta y_P)^T$. The fluctuations of the cross-quadratures are governed by the equation
\begin{eqnarray}
\partial_t \delta {\bf X} &=& \int_{-\infty}^t dt' {\bf \Sigma}_X(t-t') \delta {\bf X}(t') + {\bf v}_X(t) \\
\partial_t \delta {\bf Y} &=& \int_{-\infty}^t dt' {\bf \Sigma}_Y(t-t') \delta {\bf Y}(t') + {\bf v}_Y(t) 
\end{eqnarray}
where ${\bf \Sigma}_{X,Y} = {\bf R} {\bf M}_{\alpha, \beta} {\bf R}^T$ and ${\bf v}_{X,Y} = {\bf R}{\bf v}_{\alpha, \beta}$. By moving to the frequency domain, the above equations can be recast as
\begin{equation}
\left( \begin{array}{c} \delta \tilde{x}_+ \\ \delta \tilde{x}_- \\ \delta \tilde{x}_P \\ \delta \tilde{y}_+ \\ \delta \tilde{y}_- \\ \delta \tilde{y}_P \end{array} \right) = -({\bf \Sigma} + i \omega {\bf I})^{-1} \tilde{{\bf v}} 
\end{equation}
where $\delta \tilde{x}_+$ denotes the fourier transform of $\delta x_+$ etc. and 
\begin{equation}
{\bf \Sigma}(\omega) = \left( \begin{array}{cc} \tilde{\bf \Sigma}_X & 0 \\ 0 & \tilde{\bf \Sigma}_Y \end{array}\right)
\end{equation}
We note that in the $U(1) \times \mathbb{Z}_2$ phase where $\Delta \neq 0$, the susceptibility matrix does not remain block diagonal due to time-dependent correlations between the various cross-quadratures. However, the procedure for evaluating the eigenvalues and stability remains the same. 

The poles of the susceptibility matrix are defined by complex $\omega$ satisfying $\mathrm{Det}[{\bf \Sigma} + i \omega {\bf I}] = 0$ and the eigenvalues of the susceptibility matrix are defined as $\lambda = - i \omega$. The real part of these eigenvalues corresponds to the damping rate of the system's response to generic perturbations. As such, the stability of the mean field dynamical phases is indicated by a non-positive real part of the eigenvalues. Critical points governing continuous transitions between distinct dynamical phases is indicated by a vanishing of this real part or equivalently, a divergent relaxation time. The linearization around the trivial (parametric amplifier) solution yields the eigenvalues
\begin{equation}
\lambda_\pm = \frac{\gamma_0}{4} \left[ (\mu - \frac{2}{\gamma_0 \tau_r}) \pm \sqrt{(\mu + \frac{2}{\gamma_0 \tau_r})^2 - \frac{8}{\gamma_0 \tau_r}} \right]. 
\end{equation}
As such, for $(\gamma_0 \tau_r)^{-1} > 2$, the eigenvalues are purely real and the system can be mapped onto the Markovian system. As the coherence time of the reservoir is increased, the eigenvalues morph into complex conjugate pairs for sufficiently small drive strength $\mu$. This change from real eigenvalues to complex conjugate eigenvalues occurs at a point where the two eigenvalues (and corresponding eigenmodes) coalesce. Such points are called exceptional points. In this system, the exceptional point approaches the critical point from below as the coherence time $\tau_r$ is increased. For $(\gamma_0 \tau_r)^{-1} = \frac{1}{2}$, the exceptional point and the critical point coincide at $\mu = 1$. This heralds the emergence of the $U(1) \times \mathbb{Z}_2$ phase. For even larger reservoir coherence times, $(\gamma_0 \tau_r)^{-1} < \frac{1}{2}$, the exceptional point occurs beyond the critical point governing the transition from the disordered phase to the $U(1) \times \mathbb{Z}_2$ phase. This behavior of the exceptional point relative to the critical point is depicted in Fig. 3 of the main text. 

Similar linearization can also be performed around the $U(1)$ and the $U(1) \times \mathbb{Z}_2$ phases using the formalism described above. The calculations, while straightforward, are laborious and are not reproduced here. The real parts of the respective eigenvalues as a function of drive strength $\mu$ and normalized reservoir coherence time $(\gamma_0 \tau_r)^{-1}$ are shown in Fig. 3. As can be seen, the real parts of the eigenvalues of the susceptibility matrix for these phases are negative for large $\mu$ indicating that these dynamical phases are indeed stable to generic perturbations.

\subsection*{Two-mode correlations and entanglement}
As shown in \cite{chakram2015}, the fluctuation spectra and two-mode correlations are obtained from the power spectral densities of the cross-quadratures via the relation
\begin{equation}
{\bf S}_{X,Y}(\omega) = \frac{1}{2 \pi} (\tilde{\bf \Sigma}_{X,Y} + i \omega {\bf I})^{-1} {\bf D} (\tilde{\bf \Sigma}^\dagger_{X,Y} - i \omega {\bf I})^{-1}
\end{equation}
where the diffusion matrix is given by 
\begin{equation}
{\bf D} = \frac{1}{2} \left( \begin{array}{ccc} \frac{4 g^2}{\gamma_0 \gamma_P} \tilde{\gamma}'(\omega) (\bar{n}_{th,i} + \frac{1}{2}) & 0 & 0 \\ 0 & \frac{4 g^2}{\gamma_0 \gamma_P} \tilde{\gamma}'(\omega) (\bar{n}_{th,s} + \frac{1}{2}) & 0 \\ 0 & 0 & \frac{4 g^2}{\gamma_0^2} \gamma_P (\bar{n}_{th,P} + \frac{1}{2})  \end{array} \right)
\end{equation}
where $\tilde{\gamma}'(\omega) = \mathrm{Re}[\tilde{\gamma}(\omega)] = \gamma_0 \frac{1}{1 + (\omega \tau_r)^2}$ and the thermal phonon numbers are related to the effective temperature of the modes, i.e. $\bar{n}_{th, i,s,P} = (\exp(\frac{\hbar \omega_{i,s,P}}{k_B T}) - 1)^{-1}$. The steady state variances of the cross-quadratures can be obtained from the fluctuation spectrum using the Wiener-Khintchine theorem by integrating the fluctuations, 
\begin{equation}
{\bf \sigma}_{X,Y} = \int_{-\infty}^\infty {\bf S}_{X,Y} (\omega) d \omega
\end{equation}

Below threshold, the squeezed and amplified variances (normalized to the thermal variances) are respectively given by
\begin{eqnarray}
\sigma_{sq} &=& \frac{2 (\gamma_0 \tau_r)^{-1}}{(1 + \mu) (2 (\gamma_0 \tau_r)^{-1} + \mu)} \\
\sigma_{amp} &=&  \frac{2 (\gamma_0 \tau_r)^{-1}}{(1 - \mu) (2 (\gamma_0 \tau_r)^{-1} - \mu)}
\end{eqnarray}

Above threshold, in the $U(1)$ phase, the normalized variances of the various cross-quadratures are given by
\begin{eqnarray}
\sigma_{x_+} &=& \frac{\bar{n}_{th,P} + \frac{1}{2}}{\bar{n}_{th} + \frac{1}{2}} \,\,\frac{2 (\mu - 1)(\mu + (\gamma_0 \tau_r)^{-1})}{\mu (2 (\gamma_0 \tau_r)^{-1} + 2 \mu - 1)} + \frac{(\gamma_0 \tau_r)^{-1}}{\mu (2 (\gamma_0 \tau_r)^{-1} + 2 \mu - 1)} \\
\sigma_{x_-} &\rightarrow& \infty \\
\sigma_{y_+} &=& \frac{\bar{n}_{th,P} + \frac{1}{2}}{\bar{n}_{th} + \frac{1}{2}} \,\,\frac{2 (\mu - 1 + (\gamma_0 \tau_r)^{-1})}{ (2 (\gamma_0 \tau_r)^{-1} + 2 \mu - 3)} + \frac{(\gamma_0 \tau_r)^{-1}}{(\mu - 1) (2 (\gamma_0 \tau_r)^{-1} + 2 \mu - 3)} \\
\sigma_{y_-} &=& \frac{(\gamma_0 \tau_r)^{-1}}{1 + 2 (\gamma_0 \tau_r)^{-1}}
\end{eqnarray}
where we have assumed that $\bar{n}_{th} \equiv \bar{n}_{th,i} \approx \bar{n}_{th,s}$. 

In the $U(1) \times \mathbb{Z}_2$ phase, the non-zero frequency shifts $\Delta$ introduce correlations between the nominally uncorrelated Langevin forces in orthogonal cross-quadratures. In addition, as mentioned previously, this frequency shift also introduces time-dependent correlations between the various cross-quadratures. 
Aside from these modifications, the computation of the various variances proceeds as before. The final expressions are cumbersome and not reproduced here. 
Lastly, the logarithmic negativity is obtained from the squeezed variances as $\mathcal{E}_N = -\frac{1}{2} \log_2 \left[ \mathrm{min} (\frac{\sigma_{sq}}{\sigma_{zpm}}, 1) \right]$, where $\sigma_{zpm}$ is the zero point variance of the cross-quadratures. These results are shown in Fig. 4 of the main text.

\end{document}